\documentclass[preprint,5p,twocolumn]{elsarticle}

\usepackage{slashed}
\usepackage{graphicx}
\usepackage{amssymb,amstext,amsbsy,amsmath,amsfonts}
\usepackage{nicefrac}
\usepackage{colortbl}
\usepackage[dvipsnames]{xcolor}
\usepackage{multirow}

\usepackage{caption} 
\usepackage[colorlinks,citecolor=byzantium,linktoc=all,linkcolor=blue(ncs),urlcolor=blue(ncs)]{hyperref}

\usepackage{lineno}
\biboptions{numbers,sort&compress}

\usepackage{etoolbox}
\makeatletter
\patchcmd{\ps@pprintTitle}{\footnotesize\itshape
       Preprint submitted to \ifx\@journal\@empty Elsevier
       \else\@journal\fi}{
       \footnotesize\itshape Preprint MITP/18-122}{}{}
\makeatother

\newcommand{\beq}{\begin{equation}}
\newcommand{\eeq}{\end{equation}}
\newcommand{\bea}{\begin{eqnarray}}
\newcommand{\eea}{\end{eqnarray}}
\newcommand{\nn}{\nonumber}
\def\eqlab#1{\label{eq:#1}}
\def\figlab#1{\label{fig:#1}}

\def\barr{\left(\begin{array}{c}}
\def\earr{\end{array}\right)}
\def\bmat{\left(\begin{array}{cc}}
\def\emat{\end{array}\right)}
\def\eref#1{(\ref{eq:#1})}
\def\Eqref#1{Eq.~(\ref{eq:#1})}

\def\Figref#1{Fig.~\ref{fig:#1}}

\DeclareMathOperator{\im}{Im}
\def\nn{\nonumber}
\def\dd{\mathrm{d}}

\def\de{\delta} \def\De{\Delta}
  \def\eps{\epsilon}

\def\vfi{\varphi}

\def\dd{{\rm d}}

\begin{document}

\begin{frontmatter}

%% Title, authors and addresses

\title{Lower bound on the proton charge radius from electron scattering data}

%% use the tnoteref command within \title for footnotes;
%% use the tnotetext command for the associated footnote;
%% use the fnref command within \author or \address for footnotes;
%% use the fntext command for the associated footnote;
%% use the corref command within \author for corresponding author footnotes;
%% use the cortext command for the associated footnote;
%% use the ead command for the email address,
%% and the form \ead[url] for the home page:
%%
%% \title{Title\tnoteref{label1}}
%% \tnotetext[label1]{}
%% \author{Name\corref{cor1}\fnref{label2}}
%% \ead{email address}
%% \ead[url]{home page}
%% \fntext[label2]{}
%% \cortext[cor1]{}
%% \address{Address\fnref{label3}}
%% \fntext[label3]{}

%% use optional labels to link authors explicitly to addresses:
%% \author[label1,label2]{<author name>}
%% \address[label1]{<address>}
%% \address[label2]{<address>}

\author{Franziska Hagelstein}
\address{Albert Einstein Center for Fundamental Physics, Institute for Theoretical Physics, University of Bern, Sidlerstrasse 5, CH-3012 Bern, Switzerland}
\author{Vladimir Pascalutsa}
\address{Institut f\"ur Kernphysik and Cluster of Excellence PRISMA, Johannes Gutenberg Universit\"at Mainz,  D-55128 Mainz, Germany}

\begin{abstract}
%% Text of abstract
The proton charge-radius determinations from the 
electromagnetic form-factor measurements in 
electron-proton ($ep$) scattering require an extrapolation
to zero momentum transfer ($Q^2=0$) which is prone to model-dependent
assumptions. We show that the data at finite momentum transfer
can be used to establish a rigorous lower bound on the proton charge 
radius. Using the available $ep$ data at low $Q^2$ (below 0.02 GeV$^2$), we
obtain $R_E > 0.848$ fm (with 95\% confidence) as the lower bound on the proton radius. 
This result takes into account the statistical errors of the experiment, whereas the systematic errors are assumed to contribute to
the overall normalization of the $ep$ cross section only.
With this caveat in mind, the obtained lower bound is on the edge of reaffirming the discrepancy between the $ep$ and muonic-hydrogen
values, while bypassing the model-dependent assumptions that
go into the fitting and extrapolation of the $ep$ data.
The near-future precise $ep$ experiments at very low $Q^2$,
such as PRad, are expected to set a more stringent bound.
\end{abstract}

\begin{keyword}
Charge radius \sep proton size \sep  form factors \sep charge distribution 
\sep electron scattering

%\PACS
\end{keyword}

\end{frontmatter}

%%
%% Start line numbering here if you want
%%
%\linenumbers
\tableofcontents

%\newpage

\section{Introduction}

The proton charge radius is traditionally accessed
in elastic  electron-proton ($ep$)  scattering at small momentum transfers (low $Q$) \cite{Hofstadter:1955ae,Hofstadter:1957wk}.
Recently, however, the accuracy of this method 
has been questioned in the context  of the {\it proton-radius
puzzle}, which is partially attributed 
to the discrepancy between the 2010
$ep$ scattering value of Bernauer et al.~\cite{Bernauer_High_2010,Bernauer:2014} and the muonic-hydrogen ($\mu$H) extraction of the proton radius \cite{Pohl:2010zza,Antognini_Proton_2013}, see \Figref{R2pSummary}. 
Meanwhile, as seen from the figure,
the different extractions based on 
$ep$-scattering data have covered a whole range of values
and hardly add-up into a coherent picture. 

\begin{figure}[tbh]
\centering
	\includegraphics[width=8.5cm]{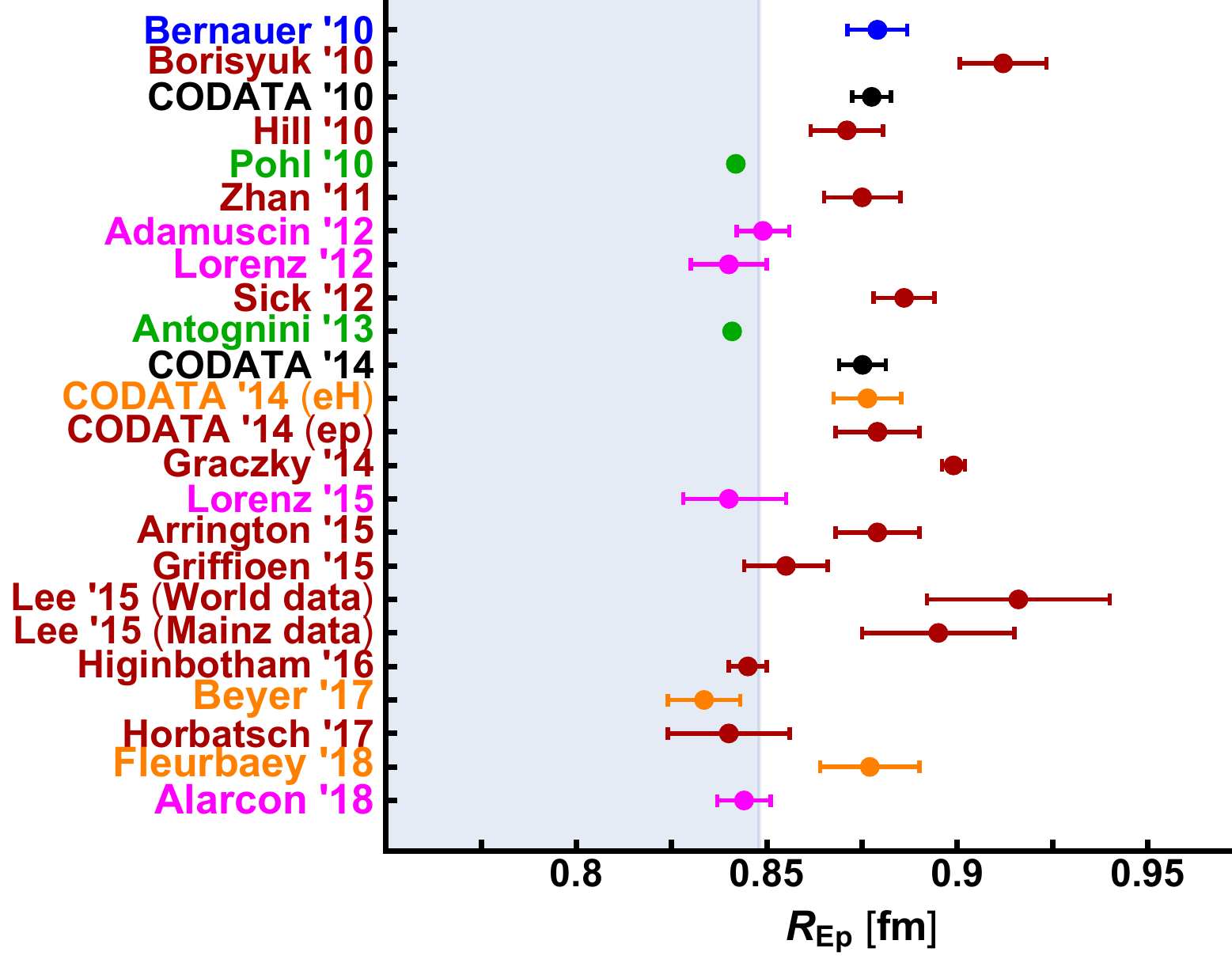}
	\caption{Summary of different proton charge-radius extractions. A) CODATA recommended charge radii in black: '10 \cite{Mohr:2012aa}, '14 \cite{Mohr_CODATA_2016}. B) hydrogen and deuterium spectroscopy in orange: Beyer '17 \cite {Beyer79}, Fleurbaey '18 \cite{Fleurbaey:2018fih}; C) muonic-hydrogen spectroscopy in green: Pohl '10 \cite{Pohl:2010zza}, Antognini '13 \cite{Antognini_Proton_2013}; D) electron-proton scattering experiments in red: Borisyuk '10 \cite{Borisyuk:2009mg}, Hill '10 \cite{Hill:2010yb} ($z$ expansion), Zhan '11 \cite{Zhan:2011ji} (recoil polarimetry), Sick '12 \cite{Sick_Problems_2012}, Graczky '14 \cite{Graczyk:2014lba}, Arrington '15 \cite{Arrington:2015ria}, Griffioen '15 \cite{Griffioen:2015hta}, Lee '15 \cite{Lee:2015jqa}, Higinbotham '16 \cite{Higinbotham:2015rja}, Horbatsch '17 \cite{Horbatsch:2016ilr} (fit with chiral perturbation theory input for higher moments); E) electron-proton scattering fits within a dispersive framework in magenta:  Adamuscin '12 \cite{Adamuscin:2012zz}, Lorenz '12 \cite{Lorenz:2012tm}, Lorenz '15 \cite{Lorenz_Theoretical_2015}, Alarcon '18 \cite{Alarcon:2018zbz};  F) electron-proton scattering data from Bernauer '10 \cite{Bernauer_High_2010} in blue. G) the values excluded by lower bound from this work are indicated by the light-blue band.}
	\figlab{R2pSummary}
\end{figure}

A ``weak link'' of the proton-radius extractions from $ep$ experiments is the extrapolation
to zero momentum transfer.  Namely,
while the data taken in some finite-$Q^2$ range can directly be  mapped into the proton (electric and magnetic) Sachs form factors
$G_E(Q^2)$ and $G_M(Q^2)$, the radii extractions require
the derivatives of those at $Q^2=0$, e.g.: $R_E =\sqrt{-6\, G_E'(0)}$. As much as one  believes that the 
slope at $0$ is largely 
determined by the behavior at finite $Q^2$,
it is not easy to quantify this relation with the necessary precision. The issues of fitting and extrapolation of the 
form-factor data have lately been under intense discussion,
see, e.g., Refs.~\cite{Sick_Problems_2012,Bernauer:2016ziz,Hayward:2018qij,Yan:2018bez}.
Similar extrapolation problems should exist in the extractions based on lattice QCD, since the lowest momentum-transfer therein is severely limited by the finite volume.

Here, we show that the form-factor data at finite $Q^2$ provide a {\it lower bound} on the proton charge radius. A determination of this bound needs no extrapolation,
therefore no major model assumptions, and should be based solely on experimental (or lattice) data. At the same time, given that some of the conventional extractions
from $ep$ data show a considerably larger radius than the $\mu$H value, a strict
lower bound, based purely on data,
is potentially useful in understanding this discrepancy.  

In what follows, we briefly recall the basic formulae in Sec.\ 2, introduce the quantity proposed to serve as the
charge-radius bound  in Sec.\ 3, obtain an empirical value for it based on proton electric form-factor data in Sec.\ 4 and conclude in Sec.\ 5.

\section{Basic ingredients of the radius extraction}

Let us recall that a spin-1/2 particle, such as the proton, has two electromagnetic form factors. These are either the Dirac and Pauli form factors: $F_1(Q^2)$
and $F_2(Q^2)$; or, the electric and magnetic Sachs form factors: 
\begin{subequations}
\bea 
\eqlab{GEdef}
G_E(Q^2)&=&
F_1(Q^2) -\frac{Q^2}{4M^2}\, F_2(Q^2),\\
G_M(Q^2)&=&F_1(Q^2)+F_2(Q^2),
\eea 
\end{subequations}
with $M$ the particle mass. The Sachs form factors can be interpreted as the Fourier transforms of the charge and magnetization
distributions, $\rho_E(\vec r\,)$ and  $\rho_M(\vec r\,)$, in the Breit frame. Strictly
speaking, this relation holds only for spherically symmetric densities, in which
case one has, see e.g.\ Ref.~\cite{Perdrisat:2006hj}:
\begin{subequations}
\bea  
\eqlab{Fdef}
G_{E}(Q^2) &=&  4\pi\int_0^\infty\! \dd r \, r^2
j_0(Qr)\,
\rho_{E}(r), \\
\frac{G_{M}(Q^2)}{1+\kappa} &=&  4\pi\int_0^\infty\! \dd r \, r^2
j_0(Qr)\,
\rho_{M}(r),
\eea 
\end{subequations}
where  $j_0(x)=\frac{\sin x}{x}$ is the spherical Bessel function, and $\kappa$ is the anomalous magnetic moment of the proton. Note that these are Lorentz-invariant expressions, hence, the spherically symmetric charge and magnetization distributions are, just as the form factors, Lorentz-invariant quantities.

The radii are introduced through the density moments,
which, for even $k$, can be given by the form-factor derivatives at $0$:
\bea 
\left<r^{k}\right>_{E} &\equiv&
4\pi \int_0^\infty\! \dd r \, r^2 \, r^{k}
\rho_{E}(r) \nn\\
&\stackrel{\mathrm{even}\, k}{=} & (-1)^{k/2}  \frac{(k+1)!}{(k/2)!} 
G_{E}^{(k/2)}(0);
\eea
and similarly for the magnetic radii with $\rho_E$ replaced by $\rho_M$, and $G_E$ replaced by $G_M/(1+\kappa)$, respectively.
Therefore, the Taylor expansion of the form factor around $Q^2=0$ is written as:
\bea
\hspace{-0.4cm}G_{E}(Q^2) &=& \sum_{n=0}^\infty  \frac{(-1)^n}{(2n+1)!}\left<r^{2n}\right>_{E} Q^{2n}\nn\\
&=& 1 - \frac{1}{6} \left<r^2\right>_E Q^2 +  \frac{1}{120} \left<r^4\right>_E Q^4 +\ldots
\eqlab{Texp}
\eea

The subject of interest  is the  
root-mean-square (rms) radius (or, simply the charge radius): $R_E = \sqrt{\left<r^{2}\right>_E}$. Ideally, it could be extracted by fitting the first few
terms of the above Taylor expansion of the 
form factor to the experimental data at low $Q^2$. In practice, however, this does not work.
The main reason is that the convergence radius of the Taylor expansion is limited
by the onset of the pion-production branch cut for time-like photon momenta at $Q^2=-4m_\pi^2$
(the nearest singularity, as far as the strong interaction is concerned), and there
are simply not many $ep$ data for $Q^2 \ll 4 m_\pi^2 \approx 0.08$ GeV$^2$.

A viable approach to fit to higher $Q^2$ is, instead of the Taylor expansion, to use a form which
takes the singularities into account. This is done in the $z$-expansion \cite{Hill:2010yb} and dispersive
fits \cite{Adamuscin:2012zz,Lorenz_Theoretical_2015,Alarcon:2018zbz}. These approaches have, however,
other severe limitations. The $z$-expansion only deals with the first singularity and therefore extends the convergence radius to $9m_\pi^2$ only. The dispersive approach
is based on an exact dispersion relation for the form factor:
\beq
\eqlab{subDR}
G_E(Q^2) = 1 - \frac{Q^2}{\pi}\int\limits_{4m_\pi^2}^\infty\! \dd t \, \frac{\im G_E(t) }{t(t+Q^2)},
\eeq 
which, in principle, accounts for all singularities. Unfortunately, it requires the knowledge
of the spectral function, $\im G_E(t)$, which is not directly accessible in experiment, and needs to be modeled. Chiral perturbation theory can only provide
a description of this function in the range of $t\ll 1$ GeV$^2$. 
Despite the recent progress in the empirical description of the 
spectral function \cite{Hoferichter:2016duk}, the problem of model dependence of the radius extraction
in the dispersive approach remains to be non-trivial.

\section{Positivity bounds}
Given the aforementioned issues in extracting the charge radius from form-factor data, we turn to
establishing a bound on the radius, rather than the radius itself. 
The advantage is that the bound will follow from the finite-$Q^2$ data alone and needs
no extrapolations or model assumptions.

To this end we consider the following quantity:
\beq 
\eqlab{R2def}
R_E^2(Q^2) \equiv  -\frac{6}{Q^2} \ln G_E(Q^2) ,
\eeq 
which in the real-photon limit yields the radius squared:
\beq
\lim_{Q^2\rightarrow 0}R_E^2(Q^2)=-6\,\frac{G_E'(Q^2)}{G_E(Q^2)}\bigg\vert_{Q^2=0} = R_E^2.
\eeq
As will be argued in Sec.\ 3, the spacelike ($Q^2\geq 0$) proton form factor is bounded from above:
\beq 
\eqlab{unity}
G_E(Q^2) \leq 1, 
\eeq 
and hence, the above log-function is positive, $R_E^2(Q^2) \geq 0$.
Furthermore, if $G_E$ falls with increasing $Q^2$ not faster than
by a power law, then $R_E^2(Q^2)$ falls as well. 
The analytic properties of $G_E$, in the absence
of zeros, are inherited by its logarithm.
The subtracted dispersion relation \eref{subDR} for the form factor
then leads to an unsubtracted one for $R_E^2$:
\beq
\eqlab{DR}
R_E^2 (Q^2) = \frac{1}{\pi}\int\limits_{4m_\pi^2}^\infty\! \dd t \, \frac{\im R^2_E(t) }{t+Q^2},  
\eeq 
where $\im R^2_E(t) = (6/t) \vfi_E(t)$, and $\vfi_E(t)\geq 0$ is the phase
defined through $G_E(t) = |G_E(t)| e^{i\vfi_E(t)}$.  
This dispersion relation shows that the function is monotonic in the spacelike region.
The latter allows one to establish
a \textit{lower bound on the radius}:
\beq
\eqlab{lowerbound}
R_E^2(Q^2) \leq R_E^2 , \quad \mbox{for}\; Q^2\geq 0.
\eeq 
Substituting in here the Taylor expansion, \Eqref{Texp}, one has:
\beq 
\eqlab{expanded}
R_E^2(Q^2) =  R_E^2 -   \left(\mbox{$\frac{1}{20}$} \left<r^4\right>_E - 
\mbox{$\frac{1}{12}$} R_E^4\right) Q^2+\mathcal{O}(Q^4), 
\eeq 
and so, in order for the bound to hold at arbitrarily low $Q^2$, the fourth and second moments must satisfy
the following inequality:\footnote{Based on \Eqref{DR}, one can claim that $R^2_E(Q^2)$ is completely monotonic, i.e.:
$ (-1)^n \dd^n R_E^2(Q^2)/\dd (Q^2)^n \geq 0 $, 
from which the lower bounds on other radii can be derived. The lowest values of the radii are given in terms of the charge radius $R_E$, and can all be obtained from Taylor-expanding the following form of the form factor: $G_E^{\mathrm{(min)}}(Q^2)
= \exp\!\big(\! - \frac16 R_E^2 Q^2\big)$. } 
\beq
\sqrt{\mbox{$\frac{3}{5}$}\left<r^4\right>_E} >  R_E^2.
\eeq 
We have checked  that 
this non-trivial hierarchical condition on the radii, which follows from 
the lower bound \Eqref{lowerbound}, is verified in existing empirical parametrizations of the proton form factor, of which the dipole form,
$G_E(Q^2)  = [1+ Q^2/(0.71\, \mathrm{GeV}^2)]^{-2}$, is the simplest one. 

The fact that $R_E^2(Q^2)$ is monotonically increasing towards $Q^2=0$ means that the best bound is obtained at lowest accessible
$Q^2$. In practice, however, it depends on the size of the experimental errors, including the uncertainty
in the overall normalization of the form factor. We discuss this in detail
in Sec.\ 4, when obtaining the empirical value of the bound from experimental data. In the rest of this section we focus on the
proof of \Eqref{unity}.

\begin{figure*}[th]
\centering
	\includegraphics[width=16.5cm]{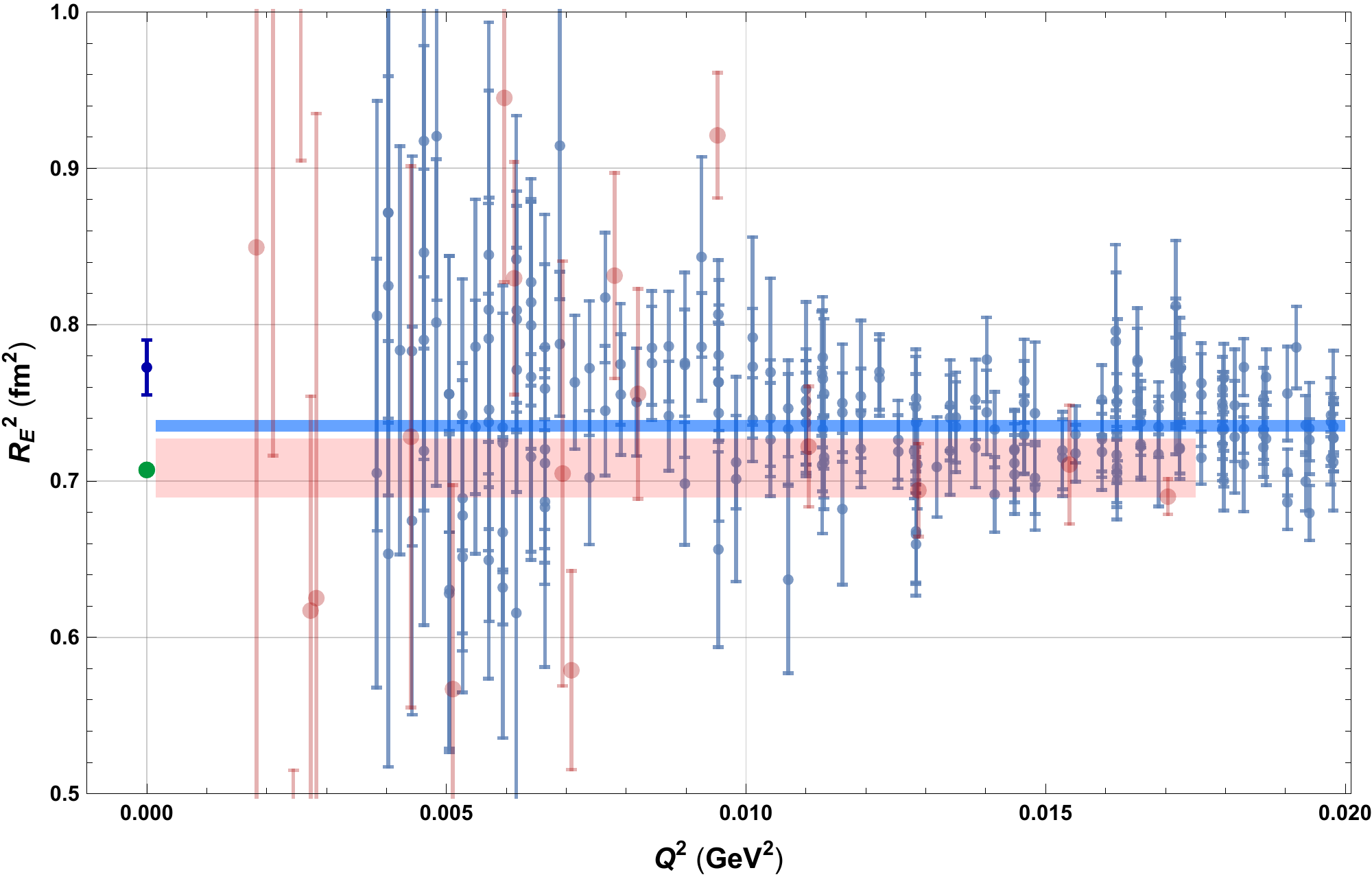}
	\caption{The quantity $R_E^2(Q^2)$ defined in \Eqref{R2def} for the proton, whose value
	at 0 represents the proton charge-radius squared. The dark-blue
	and green points at 0 indicate the $ep$ and $\mu$H values,
	respectively. The light-blue data points represent the dataset of Bernauer \textit{et al.}~\cite{Bernauer_High_2010,Bernauer:2014}.
	The light-red data points represent the ISR dataset of Mihovilovi\'c \textit{et al.} \cite{Mihovilovic:2016rkr}. The blue and red bands are the statistical averages of the corresponding datasets and are given numerically in the
	``Raw Average'' column of Table~\ref{Table1}. }
	\figlab{R2p}
\end{figure*}
\begin{figure*}[bt]
\centering
	\includegraphics[width=16.5cm]{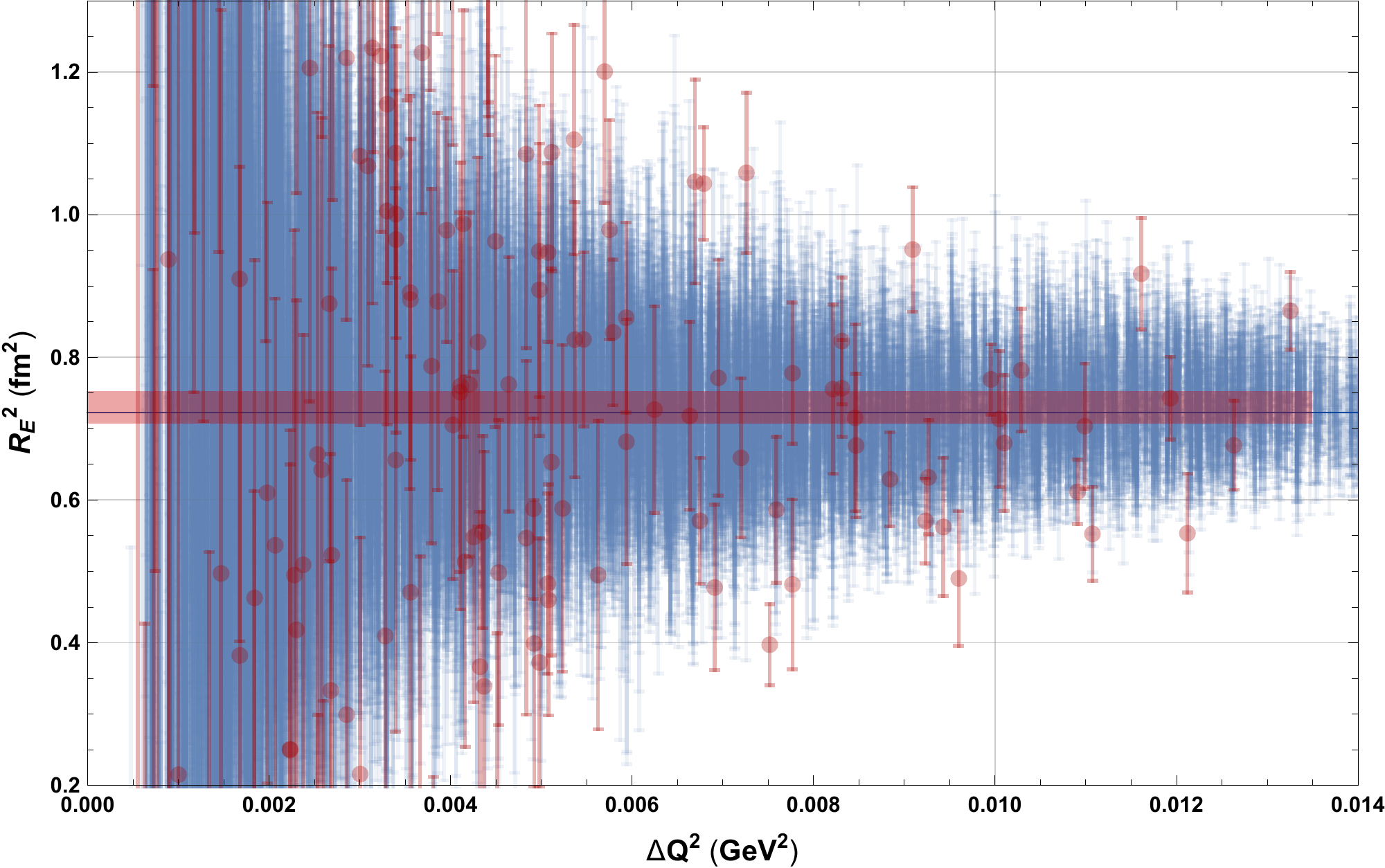}
	\caption{The quantity $R_{ij}^2$ defined in \Eqref{Rijdef} as a function of $\De Q^2 = Q_j^2-Q_i^2$ for the proton. The light-blue data points represent the dataset of Bernauer \textit{et al.}~\cite{Bernauer_High_2010}.
	The light-red data points represent the ISR dataset of Mihovilovi\'c \textit{et al.} \cite{Mihovilovic:2016rkr}. The blue and red bands show the corresponding straight-line fits, which are interpreted as a normalization-free determination of the lower bound as reflected
	in the last column of Table~\ref{Table1}. }
	\figlab{R2ij}
\end{figure*}

\begin{figure*}[bt]
\centering
	\includegraphics[width=18cm]{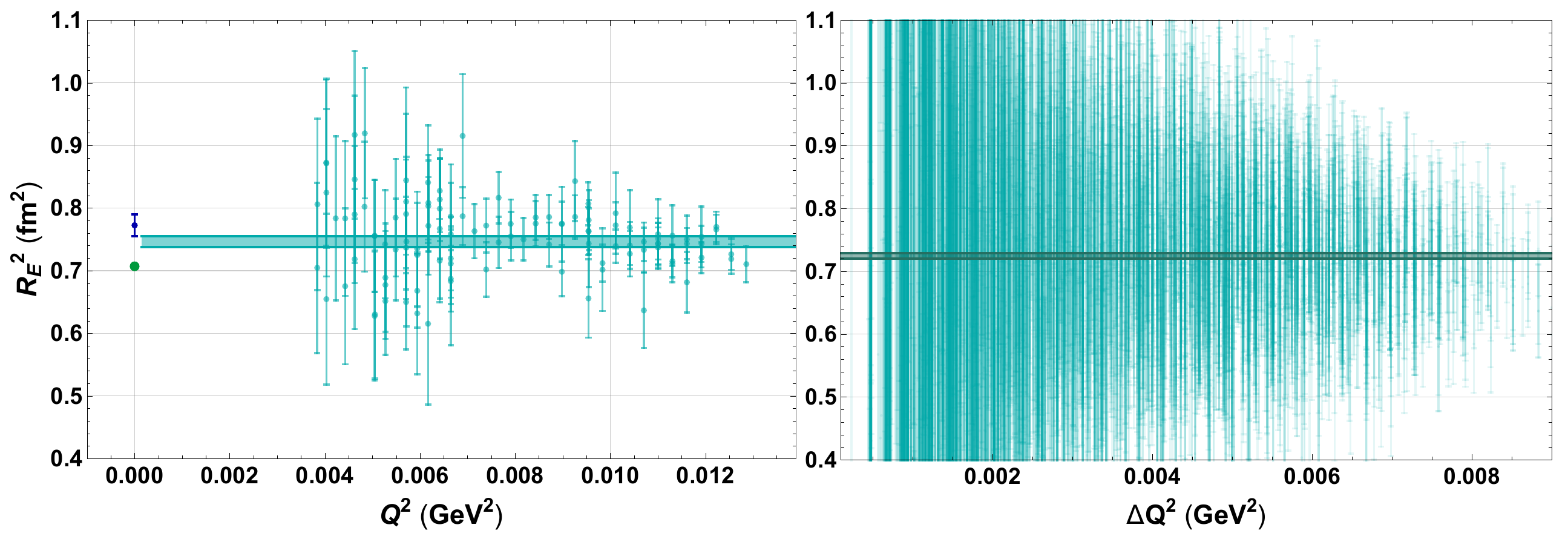}
	\caption{The quantity $R_E^2(Q^2)$ of \Eqref{R2def} (left panel) and  $R_{ij}^2(\Delta Q^2)$ of \Eqref{Rijdef} (right panel). The dark-blue
	and green points at 0 indicate the $ep$ and $\mu$H values,
	respectively. The cyan data points represent the dataset of Bernauer \textit{et al.}~\cite{Bernauer_High_2010} with normalization ``1:3''.
	 The cyan band in the left panel is the statistical averages of the corresponding dataset and is given numerically in the
	``Raw Average'' column of Table~\ref{Table1}. The green band in the right panel shows the corresponding straight-line fit, which is interpreted as a normalization-free determination of the lower bound as reflected
	in the last column of Table~\ref{Table1}. }
	\figlab{N13}
\end{figure*}

The unitary bound on the proton form factor, given in \Eqref{unity}, and subsequently
the radius bound, given in \Eqref{lowerbound}, follow from \textit{positivity} of the
corresponding charge density distribution: $\rho_E(r) \geq 0$. Indeed, from \Eqref{Fdef}, 
\beq 
\eqlab{posarg}
G_E(0) - G_E(Q^2) = 4\pi\int_0^\infty\! \dd r \, r^2
\big[1 -j_0(Qr)\big]\,
\rho_{E}(r),
\eeq 
with the property of the Bessel function $j_0(x)\leq 1$, and the positivity
of $\rho_E(r)$, we can see see that the integrand on the right-hand side 
is positive definite,  and \Eqref{unity} follows upon substituting $G_E(0)=1$
on the left-hand side.

There is a concern \cite{Miller:2010nz} that the proton charge density is not necessarily
positive definite, and only the transverse charge density is ($\rho_\perp(b)\geq 0$).
The latter relates to the Dirac form factor through the two-dimensional
Fourier transform:
\beq 
F_1(Q^2)
= 2\pi\int_0^\infty\! \dd b \, b\,
J_0(Qb)\,
\rho_\perp(b),
\eqlab{transverse}
\eeq 
where $J_0(x)$ is the cylindrical Bessel function. However,
the positivity of the transverse charge density is sufficient to
prove the unity bound of \Eqref{unity}. To see this, one may
apply the previous argument [cf.~\Eqref{posarg}] to \Eqref{transverse} using
$J_0(x)\leq 1$, and derive the bound on
the Dirac form factor:
\beq
\eqlab{F1bound}
F_1(Q^2)\leq 1.
\eeq
Then, the unitary bound on $G_E$ follows from
its definition in terms of the Dirac and Pauli form factors, see \Eqref{GEdef},
by taking into account the conditions $F_1(Q^2)\leq 1$ and $F_2(Q^2)\geq 0$. The latter is valid for the proton in at least 
the low-$Q$ region, as can be seen empirically from $F_2(0)=\kappa$, with $\kappa\simeq 1.79$
the anomalous magnetic moment of the proton.
%, or from the fact that $\rho_M(r)>\rho_E(r)$ for small distances $r$ \cite{Kelly:2002if}.

While the unity bound on $G_E$ follows from the positivity of $\rho_E(r)$, the reverse
is not necessarily true. Therefore, the proof based on the positivity of the transverse 
charge density $\rho_\perp(b)$ does
not necessarily imply the positivity of $\rho_E(r)$. Introducing $\rho_1(r)$ as the three-dimensional
Fourier-transform of the Dirac form factor, we have:
\beq  
\eqlab{F1def}
F_{1}(Q^2) =  4\pi\int_0^\infty\! \dd r \, r^2
j_0(Qr)\,
\rho_{1}(r),  
\eeq
and matching it to \Eqref{transverse}, we obtain its relation to the transverse 
density:\footnote{Here we recall the following relations between the spherical
and cylindrical Bessel functions:
\bea  
J_0(x) &=& \frac{2}{\pi} \int_x^\infty \dd x' \, x' \, 
\frac{j_0(x')}{\sqrt{x^{\prime\,2} - x^2}},\nn\\
j_0(x) &=& \frac{1}{x} \int_0^x \dd x' \, x' \, 
\frac{J_0(x')}{\sqrt{x^2-x^{\prime\,2} }},\nn
\eea
as well as their orthogonality:
\bea 
\int_0^\infty \dd Q \, Q \, J_l(Qb)\, J_l(Qb')
&=& \frac{1}{b}\de(b-b'), \nn\\
\int_0^\infty \dd Q \, Q^2 \, j_l(Qr)\, j_l(Qr')
&=& \frac{\pi}{2r^2}\de(r-r').\nn
\eea 
}
\begin{subequations}
\bea 
\rho_\perp (b) &=& 2 \int_b^\infty 
\dd r \,   
\frac{r}{\sqrt{r^{2} - b^2}}\,\rho_1(r)\\
&=&\int_{-\infty}^\infty \dd z \, \rho_1\big(\sqrt{b^2+z^2}\big) .
\eea 
\end{subequations}
The two are thus related by the \textit{Abel transform} \cite[p.\ 351 et seqq.]{Bracewell}. It infers $\rho_\perp\geq 0 $, for $\rho_1 \geq 0 $, 
while the reverse is not necessarily true.

\section{Exploring the $ep$ scattering data}

\subsection{Direct determination}
We now proceed to obtaining the lower bound on the proton charge radius from
$ep$ scattering data. The first step is to convert the experimental data for $G_E(Q^2)$
to $R_E^2(Q^2)$, using the definition \eref{R2def}. 
The presently available data 
in the region well below the pion-pair production scale (here we chose $Q^2<0.02$ GeV$^2$)
are shown in \Figref{R2p}. The light-blue points are from the
dataset of Bernauer et al.\ \cite{Bernauer_High_2010,Bernauer:2014}. The light-red data
points are from the recent initial-state radiation (ISR) experiment at MAMI \cite{Mihovilovic:2016rkr}. In both cases we deal with the statistical error bars only. 
The two points at $Q^2=0$ indicate the muonic-hydrogen (green) and 
Bernauer's $ep$-scattering (dark-blue) values of the proton charge radius.

In principle, every data point in \Figref{R2p}, at finite $Q^2$, provides a lower bound  
on the proton charge radius. For a more accurate value, we can average
over any subset of these data. In the figure, the horizontal blue band is the statistical average of  Bernauer's dataset, whereas the red band is
the statistical average of the ISR dataset. The corresponding values for the  lower bound are presented in the
``Raw-average'' column of Table~\ref{Table1}.

This is how ideally the bound should be determined from the experimental data.
However, the present experimental data have systematic uncertainties of which the most acute
one is the unknown absolute normalization of the cross section. The Bernauer dataset, for example, is 
normalized in conjunction with the radius extraction. Thus, the data normalization and the extrapolation
to $Q^2=0$ are done simultaneously in the same fit. Moreover, one can obtain an equally good
representation of Bernauer's data by using  a lower value of the radius and different normalization factors
\cite{Higinbotham:privcomm, Higinbotham:2019jzd}. 
In what follows,
we attempt to deal with this problem and construct a lower bound which is independent of the overall
normalization.

\subsection{Overall normalization factor}
To see how the normalization uncertainty affects the bound,
 let us suppose the experimental form factor has a small normalization error $\eps$,
such that $G_E^{\mathrm{(exp)}} = (1+\eps) \,G_E$, with $G_E$ having the usual interpretation.
Then,
\bea 
R_E^{2\mathrm{(exp)}}(Q^2) &=&  -\frac{6}{Q^2} \ln \big[(1+\eps)\, G_E(Q^2) ]\nn\\
&=&  R_E^2(Q^2) -\frac{6}{Q^2}\ln(1+\eps) .
\eea
If $\eps$ is positive, this is not a problem --- the lower bound
is preserved: $R_E^{2\mathrm{(exp)}}(Q^2) \leq R_E^2$, for $\eps\geq 0$. In the case of $\eps<0$, in a certain low-$Q^2$ region,
the bound is violated:
\beq 
R_E^{2\mathrm{(exp)}}(Q^2) \nleq R_E^2,\quad \mbox{for $Q^2< Q_0^2$},
\eeq 
where $Q_0$ is the root of the following equation:
\beq 
R_E^2(Q_0^2) - \frac{6}{Q_0^2}\ln (1+\eps) = R_E^2.
\eeq 
Assuming $Q_0$ is small, we can use the expanded form of $R_E^2(Q^2)$ in \Eqref{expanded}, to find:
\beq
Q_0^2 = \sqrt{\frac{-6\ln(1+\eps)}{\mbox{$\frac{1}{20}$} \left<r^4\right>_E - 
\mbox{$\frac{1}{12}$} R_E^4}}.
\eeq 
For example, taking $\eps=-0.001$ and typical values of the radii \cite{Distler:2010zq}, this equation gives $Q_0^2\approx 0.01$ GeV$^2$. 
Therefore, one strategy for avoiding the possible normalization issue is to drop the data below a certain $Q^2$ value from the 
lower-bound evaluation. A more efficient strategy is to use values at different $Q^2$ to cancel the overall normalization, as illustrated in what follows.

\renewcommand{\arraystretch}{1.5}
\begin{table*}[h]
\caption{The lower-bound value of the proton charge radius, $R_E$ (in fm), from 
two experiments and three experimental
data sets.
 The error corresponds to the 95\% confidence interval (i.e., $\pm 2\sigma$), obtained from statistical errors alone. These results are represented by the bands in Figs.~\ref{fig:R2p}, \ref{fig:R2ij} and \ref{fig:N13}
 with the corresponding color-coding. \label{Table1}}
 \centering
\begin{tabular}{|c|c|c|c|}
\hline
 \rowcolor[gray]{.95}
 \multicolumn{2}{|c|}{Dataset} &Raw Average & Normalization-free \\
\hline
&$Q^2<0.02$ GeV$^2$&\color{NavyBlue} $\boxed{\color{black} 0.857\pm 0.003} $&$\color{blue}\boxed{\color{black}0.850\pm 0.001 }$\\
\multirow{ -2}{*}{Bernauer \textit{et al.}~\cite{Bernauer:2014}}
  &subset ``1:3''&\color{cyan} $\boxed{\color{black}0.864 \pm 0.005} $&$\color{PineGreen}\boxed{\color{black}0.851\pm 0.003 }$ \\
 \rowcolor[gray]{.95} Mihovilovi\'c et al.\ \cite{Mihovilovic:2016rkr} &all data&$\color{pink}\boxed{\color{black}0.842 \pm 0.011}$ &$\color{red}\boxed{\color{black}0.854 \pm 0.014}$\\[1mm]
\hline
\end{tabular}
\end{table*}
\renewcommand{\arraystretch}{1.3}

\subsection{Towards normalization-free bound}

Having the form-factor data at a number of different  points
$Q^2_i$ (with $i=\overline{1,N}$), one may consider the following
quantity:
\beq 
\eqlab{Rijdef}
R^2_{ij} \equiv \frac{-6}{Q_j^2 -Q_i^2} \ln \frac{G_E(Q_j^2)}{G_E(Q_i^2)}. 
\eeq 
The obvious advantage of this form is that the overall-normalization uncertainty
cancels out. At the same time, each element of the symmetric matrix $R_{ij}^2$ provides a lower bound:
$R^2_{ij} < R_E^2$, for any $i,j$.
This can be seen by rewriting it identically as:
\bea
\eqlab{Rij}
R_{ij}^2 &=&  R^2_E(Q_i^2)  +   Q_j^2\, \frac{R^2_E(Q_j^2)-R^2_E(Q_i^2)}{Q_j^2-Q_j^2} ,
\eea 
where $R^2_E(Q^2)$ is the lower-bound function of \Eqref{R2def}.
The second term is negative-definite, given that 
$R^2_E(Q^2)$ is monotonically decreasing, and  hence: 
\beq R_{ij}^2 <  R^2_E(Q_i^2) < R_E^2.
\eeq
Because of the first inequality, the bound obtained from  $R_{ij}^2$
is lower than the one obtained from $R^2_E(Q^2)$ and therefore is less optimal.
Yet, it may be more precise when applied to real data, because of cancellation
of systematic uncertainties which affect the absolute normalization of the experimental
cross sections. 

To illustrate the workings of this method, let us consider
\Figref{R2ij}, where we plot the elements $R_{ij}$ for the two datasets, as a
function of $\De Q^2 = Q_j^2-Q_i^2$. The blue and red bands provide
the two corresponding bounds obtained by fitting a horizontal line using the \textsc{\small NonlinearModelFit} routine of \textsc{Mathematica} \cite{Mathematica}.
The corresponding values are given in the ``Normalization-free'' column of Table~\ref{Table1}. 

Note that the error on $R_{ij}^2$ is decreasing with the increase of $\De Q^2$, and hence the obtained bounds are driven by the higher $\De Q^2$ interval. In fact, one
can apply a cut on the lowest $\De Q^2$ values, without affecting the result.

Of course, this method only works if
 all the data points of a given dataset have the same normalization
factor. In reality, the experiment of Bernauer et al.~\cite{Bernauer:2014} has a complicated normalization procedure, involving 31 normalization factors, and one can manage to obtain significant shifts of the data points 
by a different fit of these factors \cite{Higinbotham:privcomm, Higinbotham:2019jzd}. These shifts could then be considered as a systematic normalization
uncertainty which is only partially attributed to an overall normalization.

Nonetheless, one can identify subsets where the difference in normalization is overall. 
In the experimental data of Bernauer et al.\ these are, for example, 
normalization sets (see Supplement in \cite{Bernauer:2014}): 
\begin{itemize}
    \item 3 (spectrometer A, $180$ MeV beam energy),
    \item 1:3 (spectrometer B, $180$ MeV beam energy), 
    \item 6:9 (spectrometer B, $315$ MeV beam energy).
\end{itemize}
We have applied the $R_{ij}^2$ method to each of this subsets separately (for $Q^2\leq 0.02$ GeV$^2$)
and obtained
the same results (within statistical errors). The most precise result is the one from the 1:3 subset, because
it is the largest in this region. The results for this dataset are shown in  
\Figref{N13} and the second row of Table~\ref{Table1}. The latter value is indeed a normalization-free
bound.

We obtained the same results with the two datasets of Higinbotham \cite{Higinbotham:privcomm},
generated from the data of Bernauer et al.~\cite{Bernauer:2014}, and corresponding
to significantly different radii.  A subset of 106 data points in the very low-$Q$ region ($Q^2\leq 0.012$ GeV$^2$) differs in an overall factor between 
the two datasets. Applying our method to this subset leads to the lower-bound value of
$ 0.851(3)$ fm, for both datasets, in exact agreement with our normalization-free result for the subset 1:3.
We hence conclude that our method leads to a robust and accurate determination
of the lower bound on $R_E$ from the form-factor data, even when they are prone to normalization uncertainties.

The lower bound resulting from the Bernauer dataset [$R_E> 0.848$ fm at 95\% confidence level (CL)] is very accurate, although we emphasize that the error is only statistical. It can be compared with the recent proton-radius extractions in \Figref{R2pSummary}. It is somewhat in conflict with the $\mu$H values \cite{Pohl:2010zza,Antognini_Proton_2013}, and the Garching measurement of the $2S-4P$ transition frequency in H \cite{Beyer79}.

\section{Conclusion}

An extraction of the proton charge radius 
from $ep$ scattering requires an extrapolation to zero momentum transfer, which nowadays is entangled in the
analysis of $ep$ data.
We aim here to leave the extrapolation issues out of the interpretation of $ep$ data.
We show that the $ep$ scattering
may directly provide
a lower bound on the proton charge radius,
cf.~\Eqref{lowerbound} with \Eqref{R2def}.
Thus, the lower bound is a directly observable quantity 
(to the extent that the form factor is), and is
a more rigorous experimental outcome than
the charge radius itself.   

We have attempted a first determination of the lower bound
on the proton charge radius 
from the available data in the region of
$Q^2$ below 0.02 GeV$^2$. 
Our results for the two presently available experiments
are given in Table~\ref{Table1}. The last column therein shows the lower-bound
values with the overall-normalization uncertainty being
canceled out. 

The lower bound, $R_{E}> 0.848$ fm (95\% CL),  resulting from
our ``normalization-free'' analysis of the $ep$ data of {Bernauer \textit{et al.}~\cite{Bernauer:2014}, 
rules out the shaded area in \Figref{R2pSummary}.
The figure also shows the results of recent proton-radius determinations. In particular, this $ep$ bound is in disagreement with the muonic-hydrogen values (green dots). 
We emphasize that the present determination of the lower bound does not involve any fitting of the $Q^2$-dependence
with subsequent extrapolation to $Q^2=0$.
On the other hand, the present analysis
does not account for systematic errors in the experimental data, except for those that contribute to the overall normalization. 

As the lower-bound function, defined in \Eqref{R2def},
is monotonically increasing with decreasing $Q^2$, the
most stringent bound will be obtained from the lower
$Q^2$ range, provided that the accuracy does not
deteriorate with decreasing $Q^2$.
Therefore, with the forthcoming results of the PRad experiment \cite{Gasparian:2014rna,Peng:2016szv}, one hopes to obtain a much better determination of the lower bound. The PRad data will reach down to $2\times10^{-4}\, \mathrm{GeV}^2$
and include a simultaneous measurement 
of the  M{\o}ller scattering. The latter will allow to further reduce the
systematic uncertainties.

\section*{Acknowledgements}
\addcontentsline{toc}{section}{Acknowledgements}
We are grateful to Jan Bernauer, Michael Distler, Miha Mihovilovi\'c, and
Thomas Walcher for sharing their data with us and helpful communications; 
to Douglas Higinbotham for checking some of our results and an interesting
discussion; to Patricia Bickert, Ashot Gasparyan, Vadim Lensky, and Stefan Scherer for useful remarks on the manuscript.
This work was supported by the Swiss National Science Foundation and the Deutsche Forschungsgemeinschaft (DFG) through the Collaborative Research Center 1044 [The Low-Energy Frontier of the Standard Model].

%\newpage
%\section*{References}
%\addcontentsline{toc}{section}{References}

%% References
%%
%% Following citation commands can be used in the body text:
%% Usage of \cite is as follows:
%%   \cite{key}          ==>>  [#]
%%   \cite[chap. 2]{key} ==>>  [#, chap. 2]
%%   \citet{key}         ==>>  Author [#]

%% References with bibTeX database:

\bibliographystyle{model1-num-names}

%\bibliography{readcube_lowerBound}
\bibliography{lowerBound}

\begin{thebibliography}{38}
\expandafter\ifx\csname natexlab\endcsname\relax\def\natexlab#1{#1}\fi
\providecommand{\bibinfo}[2]{#2}
\ifx\xfnm\relax \def\xfnm[#1]{\unskip,\space#1}\fi
%Type = Article
\bibitem[{Hofstadter and McAllister(1955)}]{Hofstadter:1955ae}
\bibinfo{author}{R.~Hofstadter}, \bibinfo{author}{R.~W. McAllister},
\newblock \bibinfo{title}{{Electron scattering from the proton}},
\newblock \bibinfo{journal}{Phys. Rev.} \bibinfo{volume}{98}
  (\bibinfo{year}{1955}) \bibinfo{pages}{217--218}.
%Type = Article
\bibitem[{Hofstadter(1957)}]{Hofstadter:1957wk}
\bibinfo{author}{R.~Hofstadter},
\newblock \bibinfo{title}{{Nuclear and nucleon scattering of high-energy
  electrons}},
\newblock \bibinfo{journal}{Ann. Rev. Nucl. Part. Sci.} \bibinfo{volume}{7}
  (\bibinfo{year}{1957}) \bibinfo{pages}{231--316}.
%Type = Article
\bibitem[{Bernauer et~al.(2010)}]{Bernauer_High_2010}
\bibinfo{author}{J.~Bernauer}, et~al.,
\newblock \bibinfo{title}{{High-Precision} determination of the electric and
  magnetic form factors of the proton},
\newblock \bibinfo{journal}{Phys. Rev. Lett.} \bibinfo{volume}{105}
  (\bibinfo{year}{2010}) \bibinfo{pages}{242001}.
%Type = Article
\bibitem[{Bernauer et~al.(2014)Bernauer, Distler, Friedrich, and
  Walcher}]{Bernauer:2014}
\bibinfo{author}{J.~C. Bernauer}, \bibinfo{author}{M.~O. Distler},
  \bibinfo{author}{J.~Friedrich}, \bibinfo{author}{T.~Walcher},
\newblock \bibinfo{title}{Electric and magnetic form factors of the proton},
\newblock \bibinfo{journal}{Phys. Rev. C} \bibinfo{volume}{90}
  (\bibinfo{year}{2014}) \bibinfo{pages}{015206}.
%Type = Article
\bibitem[{Pohl et~al.(2010)}]{Pohl:2010zza}
\bibinfo{author}{R.~Pohl}, et~al.,
\newblock \bibinfo{title}{{The size of the proton}},
\newblock \bibinfo{journal}{Nature} \bibinfo{volume}{466}
  (\bibinfo{year}{2010}) \bibinfo{pages}{213--216}.
%Type = Article
\bibitem[{Antognini et~al.(2013)}]{Antognini_Proton_2013}
\bibinfo{author}{A.~Antognini}, et~al.,
\newblock \bibinfo{title}{Proton structure from the measurement of {2S-2P}
  transition frequencies of muonic hydrogen},
\newblock \bibinfo{journal}{Science} \bibinfo{volume}{339}
  (\bibinfo{year}{2013}) \bibinfo{pages}{417--420}.
%Type = Article
\bibitem[{Mohr et~al.(2012)Mohr, Taylor, and Newell}]{Mohr:2012aa}
\bibinfo{author}{P.~J. Mohr}, \bibinfo{author}{B.~N. Taylor},
  \bibinfo{author}{D.~B. Newell},
\newblock \bibinfo{title}{Codata recommended values of the fundamental physical
  constants: 2010},
\newblock \bibinfo{journal}{Rev. Mod. Phys.} \bibinfo{volume}{84}
  (\bibinfo{year}{2012}) \bibinfo{pages}{1527--1605}.
%Type = Article
\bibitem[{Mohr et~al.(2016)Mohr, Newell, and Taylor}]{Mohr_CODATA_2016}
\bibinfo{author}{P.~J. Mohr}, \bibinfo{author}{D.~B. Newell},
  \bibinfo{author}{B.~N. Taylor},
\newblock \bibinfo{title}{{CODATA} recommended values of the fundamental
  physical constants: 2014},
\newblock \bibinfo{journal}{Rev. Mod. Phys.} \bibinfo{volume}{88}
  (\bibinfo{year}{2016}) \bibinfo{pages}{035009}.
%Type = Article
\bibitem[{Beyer et~al.(2017)Beyer, Maisenbacher, Matveev, Pohl, Khabarova,
  Grinin, Lamour, Yost, H{\"a}nsch, Kolachevsky, and Udem}]{Beyer79}
\bibinfo{author}{A.~Beyer}, \bibinfo{author}{L.~Maisenbacher},
  \bibinfo{author}{A.~Matveev}, \bibinfo{author}{R.~Pohl},
  \bibinfo{author}{K.~Khabarova}, \bibinfo{author}{A.~Grinin},
  \bibinfo{author}{T.~Lamour}, \bibinfo{author}{D.~C. Yost},
  \bibinfo{author}{T.~W. H{\"a}nsch}, \bibinfo{author}{N.~Kolachevsky},
  \bibinfo{author}{T.~Udem},
\newblock \bibinfo{title}{The rydberg constant and proton size from atomic
  hydrogen},
\newblock \bibinfo{journal}{Science} \bibinfo{volume}{358}
  (\bibinfo{year}{2017}) \bibinfo{pages}{79--85}.
%Type = Article
\bibitem[{Fleurbaey et~al.(2018)Fleurbaey, Galtier, Thomas, Bonnaud, Julien,
  Biraben, Nez, Abgrall, and Guna}]{Fleurbaey:2018fih}
\bibinfo{author}{H.~Fleurbaey}, \bibinfo{author}{S.~Galtier},
  \bibinfo{author}{S.~Thomas}, \bibinfo{author}{M.~Bonnaud},
  \bibinfo{author}{L.~Julien}, \bibinfo{author}{F.~Biraben},
  \bibinfo{author}{F.~Nez}, \bibinfo{author}{M.~Abgrall},
  \bibinfo{author}{J.~Guna},
\newblock \bibinfo{title}{{New Measurement of the $1S-3S$ Transition Frequency
  of Hydrogen: Contribution to the Proton Charge Radius Puzzle}},
\newblock \bibinfo{journal}{Phys. Rev. Lett.} \bibinfo{volume}{120}
  (\bibinfo{year}{2018}) \bibinfo{pages}{183001}.
%Type = Article
\bibitem[{Borisyuk(2010)}]{Borisyuk:2009mg}
\bibinfo{author}{D.~Borisyuk},
\newblock \bibinfo{title}{{Proton charge and magnetic rms radii from the
  elastic $ep$ scattering data}},
\newblock \bibinfo{journal}{Nucl. Phys.} \bibinfo{volume}{A 843}
  (\bibinfo{year}{2010}) \bibinfo{pages}{59--67}.
%Type = Article
\bibitem[{Hill and Paz(2010)}]{Hill:2010yb}
\bibinfo{author}{R.~J. Hill}, \bibinfo{author}{G.~Paz},
\newblock \bibinfo{title}{{Model independent extraction of the proton charge
  radius from electron scattering}},
\newblock \bibinfo{journal}{Phys. Rev.} \bibinfo{volume}{D 82}
  (\bibinfo{year}{2010}) \bibinfo{pages}{113005}.
%Type = Article
\bibitem[{Zhan et~al.(2011)Zhan, Allada, Armstrong, Arrington
  et~al.}]{Zhan:2011ji}
\bibinfo{author}{X.~Zhan}, \bibinfo{author}{K.~Allada},
  \bibinfo{author}{D.~Armstrong}, \bibinfo{author}{J.~Arrington}, et~al.,
\newblock \bibinfo{title}{{High Precision Measurement of the Proton Elastic
  Form Factor Ratio $\mu_pG_E/G_M$ at low $Q^2$}},
\newblock \bibinfo{journal}{Phys. Lett.} \bibinfo{volume}{B 705}
  (\bibinfo{year}{2011}) \bibinfo{pages}{59--64}.
%Type = Article
\bibitem[{Sick(2012)}]{Sick_Problems_2012}
\bibinfo{author}{I.~Sick},
\newblock \bibinfo{title}{Problems with proton radii},
\newblock \bibinfo{journal}{Prog. Part. Nucl. Phys.} \bibinfo{volume}{67}
  (\bibinfo{year}{2012}) \bibinfo{pages}{473--478}.
%Type = Article
\bibitem[{Graczyk and Juszczak(2014)}]{Graczyk:2014lba}
\bibinfo{author}{K.~M. Graczyk}, \bibinfo{author}{C.~Juszczak},
\newblock \bibinfo{title}{{Proton radius from Bayesian inference}},
\newblock \bibinfo{journal}{Phys. Rev.} \bibinfo{volume}{C 90}
  (\bibinfo{year}{2014}) \bibinfo{pages}{054334}.
%Type = Article
\bibitem[{Arrington and Sick(2015)}]{Arrington:2015ria}
\bibinfo{author}{J.~Arrington}, \bibinfo{author}{I.~Sick},
\newblock \bibinfo{title}{{Evaluation of the proton charge radius from e-p
  scattering}},
\newblock \bibinfo{journal}{J. Phys. Chem. Ref. Data} \bibinfo{volume}{44}
  (\bibinfo{year}{2015}) \bibinfo{pages}{031204}.
%Type = Article
\bibitem[{Griffioen et~al.(2016)Griffioen, Carlson, and
  Maddox}]{Griffioen:2015hta}
\bibinfo{author}{K.~Griffioen}, \bibinfo{author}{C.~Carlson},
  \bibinfo{author}{S.~Maddox},
\newblock \bibinfo{title}{{Consistency of electron scattering data with a small
  proton radius}},
\newblock \bibinfo{journal}{Phys. Rev.} \bibinfo{volume}{C 93}
  (\bibinfo{year}{2016}) \bibinfo{pages}{065207}.
%Type = Article
\bibitem[{Lee et~al.(2015)Lee, Arrington, and Hill}]{Lee:2015jqa}
\bibinfo{author}{G.~Lee}, \bibinfo{author}{J.~R. Arrington},
  \bibinfo{author}{R.~J. Hill},
\newblock \bibinfo{title}{{Extraction of the proton radius from electron-proton
  scattering data}},
\newblock \bibinfo{journal}{Phys. Rev.} \bibinfo{volume}{D 92}
  (\bibinfo{year}{2015}) \bibinfo{pages}{013013}.
%Type = Article
\bibitem[{Higinbotham et~al.(2016)Higinbotham, Kabir, Lin, Meekins, Norum, and
  Sawatzky}]{Higinbotham:2015rja}
\bibinfo{author}{D.~W. Higinbotham}, \bibinfo{author}{A.~A. Kabir},
  \bibinfo{author}{V.~Lin}, \bibinfo{author}{D.~Meekins},
  \bibinfo{author}{B.~Norum}, \bibinfo{author}{B.~Sawatzky},
\newblock \bibinfo{title}{{Proton radius from electron scattering data}},
\newblock \bibinfo{journal}{Phys. Rev.} \bibinfo{volume}{C 93}
  (\bibinfo{year}{2016}) \bibinfo{pages}{055207}.
%Type = Article
\bibitem[{Horbatsch et~al.(2017)Horbatsch, Hessels, and
  Pineda}]{Horbatsch:2016ilr}
\bibinfo{author}{M.~Horbatsch}, \bibinfo{author}{E.~A. Hessels},
  \bibinfo{author}{A.~Pineda},
\newblock \bibinfo{title}{{Proton radius from electron-proton scattering and
  chiral perturbation theory}},
\newblock \bibinfo{journal}{Phys. Rev.} \bibinfo{volume}{C 95}
  (\bibinfo{year}{2017}) \bibinfo{pages}{035203}.
%Type = Article
\bibitem[{Adamuscin et~al.(2012)Adamuscin, Dubnicka, and
  Dubnickova}]{Adamuscin:2012zz}
\bibinfo{author}{C.~Adamuscin}, \bibinfo{author}{S.~Dubnicka},
  \bibinfo{author}{A.~Dubnickova},
\newblock \bibinfo{title}{{New value of the proton charge root mean square
  radius}},
\newblock \bibinfo{journal}{Prog. Part. Nucl. Phys.} \bibinfo{volume}{67}
  (\bibinfo{year}{2012}) \bibinfo{pages}{479--485}.
%Type = Article
\bibitem[{Lorenz et~al.(2012)Lorenz, Hammer, and Meissner}]{Lorenz:2012tm}
\bibinfo{author}{I.~T. Lorenz}, \bibinfo{author}{H.~W. Hammer},
  \bibinfo{author}{U.-G. Meissner},
\newblock \bibinfo{title}{{The size of the proton - closing in on the radius
  puzzle}},
\newblock \bibinfo{journal}{Eur. Phys. J.} \bibinfo{volume}{A 48}
  (\bibinfo{year}{2012}) \bibinfo{pages}{151}.
%Type = Article
\bibitem[{Lorenz et~al.(2015)Lorenz, Mei{\ss}ner, Hammer, and
  Dong}]{Lorenz_Theoretical_2015}
\bibinfo{author}{I.~Lorenz}, \bibinfo{author}{U.~Mei{\ss}ner},
  \bibinfo{author}{H.~Hammer}, \bibinfo{author}{Y.~Dong},
\newblock \bibinfo{title}{Theoretical constraints and systematic effects in the
  determination of the proton form factors},
\newblock \bibinfo{journal}{Phys. Rev. D} \bibinfo{volume}{91}
  (\bibinfo{year}{2015}) \bibinfo{pages}{014023}.
%Type = Article
\bibitem[{Alarc\'on et~al.(2018)Alarc\'on, Higinbotham, Weiss, and
  Ye}]{Alarcon:2018zbz}
\bibinfo{author}{J.~M. Alarc\'on}, \bibinfo{author}{D.~Higinbotham},
  \bibinfo{author}{C.~Weiss}, \bibinfo{author}{Z.~Ye},
\newblock \bibinfo{title}{{Proton charge radius extraction from electron
  scattering data using dispersively improved chiral effective field theory}},
\newblock \bibinfo{journal}{hep-ph/1809.06373}  (\bibinfo{year}{2018}).
%Type = Article
\bibitem[{Bernauer(2016)}]{Bernauer:2016ziz}
\bibinfo{author}{J.~C. Bernauer},
\newblock \bibinfo{title}{{Avoiding common pitfalls and misconceptions in
  extractions of the proton radius}},
\newblock \bibinfo{journal}{nucl-th/1606.02159}  (\bibinfo{year}{2016}).
%Type = Article
\bibitem[{Hayward and Griffioen(2018)}]{Hayward:2018qij}
\bibinfo{author}{T.~B. Hayward}, \bibinfo{author}{K.~A. Griffioen},
\newblock \bibinfo{title}{{Evaluation of low-$Q^2$ fits to $ep$ and $ed$
  elastic scattering data}},
\newblock \bibinfo{journal}{nucl-ex/1804.09150}  (\bibinfo{year}{2018}).
%Type = Article
\bibitem[{Yan et~al.(2018)Yan, Higinbotham, Dutta, Gao, Gasparian, Khandaker,
  Liyanage, Pasyuk, Peng, and Xiong}]{Yan:2018bez}
\bibinfo{author}{X.~Yan}, \bibinfo{author}{D.~W. Higinbotham},
  \bibinfo{author}{D.~Dutta}, \bibinfo{author}{H.~Gao},
  \bibinfo{author}{A.~Gasparian}, \bibinfo{author}{M.~A. Khandaker},
  \bibinfo{author}{N.~Liyanage}, \bibinfo{author}{E.~Pasyuk},
  \bibinfo{author}{C.~Peng}, \bibinfo{author}{W.~Xiong},
\newblock \bibinfo{title}{{Robust extraction of the proton charge radius from
  electron-proton scattering data}},
\newblock \bibinfo{journal}{Phys. Rev.} \bibinfo{volume}{C 98}
  (\bibinfo{year}{2018}) \bibinfo{pages}{025204}.
%Type = Article
\bibitem[{Perdrisat et~al.(2007)Perdrisat, Punjabi, and
  Vanderhaeghen}]{Perdrisat:2006hj}
\bibinfo{author}{C.~F. Perdrisat}, \bibinfo{author}{V.~Punjabi},
  \bibinfo{author}{M.~Vanderhaeghen},
\newblock \bibinfo{title}{{Nucleon Electromagnetic Form Factors}},
\newblock \bibinfo{journal}{Prog. Part. Nucl. Phys.} \bibinfo{volume}{59}
  (\bibinfo{year}{2007}) \bibinfo{pages}{694--764}.
%Type = Article
\bibitem[{Hoferichter et~al.(2016)Hoferichter, Kubis, Ruiz~de Elvira, Hammer,
  and Meißner}]{Hoferichter:2016duk}
\bibinfo{author}{M.~Hoferichter}, \bibinfo{author}{B.~Kubis},
  \bibinfo{author}{J.~Ruiz~de Elvira}, \bibinfo{author}{H.~W. Hammer},
  \bibinfo{author}{U.~G. Meißner},
\newblock \bibinfo{title}{{On the $\pi\pi$ continuum in the nucleon form
  factors and the proton radius puzzle}},
\newblock \bibinfo{journal}{Eur. Phys. J.} \bibinfo{volume}{A 52}
  (\bibinfo{year}{2016}) \bibinfo{pages}{331}.
%Type = Article
\bibitem[{Mihovilovi\'c et~al.(2017)}]{Mihovilovic:2016rkr}
\bibinfo{author}{M.~Mihovilovi\'c}, et~al.,
\newblock \bibinfo{title}{{First measurement of proton's charge form factor at
  very low $Q^2$ with initial state radiation}},
\newblock \bibinfo{journal}{Phys. Lett.} \bibinfo{volume}{B 771}
  (\bibinfo{year}{2017}) \bibinfo{pages}{194--198}.
%Type = Article
\bibitem[{Miller(2010)}]{Miller:2010nz}
\bibinfo{author}{G.~A. Miller},
\newblock \bibinfo{title}{{Transverse Charge Densities}},
\newblock \bibinfo{journal}{Ann. Rev. Nucl. Part. Sci.} \bibinfo{volume}{60}
  (\bibinfo{year}{2010}) \bibinfo{pages}{1--25}.
%Type = Book
\bibitem[{Bracewell(2000)}]{Bracewell}
\bibinfo{author}{R.~N. Bracewell}, \bibinfo{title}{The Fourier transform and
  its applications}, \bibinfo{publisher}{McGraw-Hill International Editions},
  \bibinfo{year}{2000}.
%Type = Article
\bibitem[{Higinbotham(2018)}]{Higinbotham:privcomm}
\bibinfo{author}{D.~W. Higinbotham},
\newblock \bibinfo{journal}{private communication}  (\bibinfo{year}{2018}).
%Type = Article
\bibitem[{Higinbotham and McClellan(2019)}]{Higinbotham:2019jzd}
\bibinfo{author}{D.~W. Higinbotham}, \bibinfo{author}{R.~E. McClellan},
\newblock \bibinfo{title}{{How Variation in Analytic Choices Can Affect
  Normalization Parameters and Proton Radius Extractions From Electron
  Scattering Data}},
\newblock \bibinfo{journal}{physics.data-an/1902.08185}
  (\bibinfo{year}{2019}).
%Type = Article
\bibitem[{Distler et~al.(2011)Distler, Bernauer, and Walcher}]{Distler:2010zq}
\bibinfo{author}{M.~O. Distler}, \bibinfo{author}{J.~C. Bernauer},
  \bibinfo{author}{T.~Walcher},
\newblock \bibinfo{title}{{The RMS Charge Radius of the Proton and Zemach
  Moments}},
\newblock \bibinfo{journal}{Phys. Lett.} \bibinfo{volume}{B 696}
  (\bibinfo{year}{2011}) \bibinfo{pages}{343--347}.
%Type = Misc
\bibitem[{Inc.(2018)}]{Mathematica}
\bibinfo{author}{W.~R. Inc.}, \bibinfo{title}{Mathematica, {V}ersion 11.3},
  \bibinfo{year}{2018}. \bibinfo{note}{Champaign, IL}.
%Type = Article
\bibitem[{Gasparian for the PRad~at
  JLab~Collaboration(2014)}]{Gasparian:2014rna}
\bibinfo{author}{A.~Gasparian for the PRad~at JLab~Collaboration},
\newblock \bibinfo{title}{{The PRad experiment and the proton radius puzzle}},
\newblock \bibinfo{journal}{EPJ Web Conf.} \bibinfo{volume}{73}
  (\bibinfo{year}{2014}) \bibinfo{pages}{07006}.
%Type = Article
\bibitem[{Peng and Gao(2016)}]{Peng:2016szv}
\bibinfo{author}{C.~Peng}, \bibinfo{author}{H.~Gao},
\newblock \bibinfo{title}{{Proton Charge Radius (PRad) Experiment at Jefferson
  Lab}},
\newblock \bibinfo{journal}{EPJ Web Conf.} \bibinfo{volume}{113}
  (\bibinfo{year}{2016}) \bibinfo{pages}{03007}.

\end{thebibliography}

\end{document}